\documentclass[runningheads]{llncs}
\usepackage[T1]{fontenc}
\usepackage{newtxtext}
\usepackage[varvw]{newtxmath}
\usepackage{graphicx}
\usepackage{amsmath}

\usepackage{amssymb}
\usepackage{algorithm}
\usepackage{algpseudocode}
\usepackage{placeins}
\usepackage{xcolor}
\usepackage{array}
\makeatletter
\AtBeginDocument{\@ifundefined{doi}{\newcommand{\doi}[1]{}}{\renewcommand{\doi}[1]{}}}
\makeatother

\usepackage[colorlinks=true,linkcolor=black,citecolor=black,urlcolor=black]{hyperref}

\makeatletter
\AtBeginDocument{%
  \def\@bibitem#1{%
    \item
    \Hy@raisedlink{\hyper@anchorstart{cite.#1}\hyper@anchorend}%
    \if@filesw
      \immediate\write\@auxout{\string\bibcite{#1}{\the\value{\@listctr}}}%
    \fi
    \ignorespaces}%
  \def\@lbibitem[#1]#2{%
    \item[{[#1]}\hfill]%
    \Hy@raisedlink{\hyper@anchorstart{cite.#2}\hyper@anchorend}%
    \if@filesw
      {\let\protect\noexpand\immediate
       \write\@auxout{\string\bibcite{#2}{#1}}}%
    \fi
    \ignorespaces}%
}
\makeatother
\begin{document}
\title{DYNAS: A Dynamic Network Penetration Testing Framework Using Deep Hierarchical Reinforcement Learning}
\titlerunning{DYNAS: Dynamic Network Penetration Testing Using DHRL}
\author{Md. Rakibul Hassan\inst{1} \and Majedul Islam\inst{1} \and Md. Ashraful Alam\inst{1} \and Muhammad Iqbal Hossain\inst{1} \and Swakkhar Shatabda\inst{1}\\[0.6em]\textit{Accepted at the NSS 2026: 20th International Conference on Network and System Security (submission version)}}
\authorrunning{M. R. Hassan et al.}
\institute{BRAC University, Bangladesh\\
\email{md.rakibul.hassan@g.bracu.ac.bd}, \email{majedul.islam@g.bracu.ac.bd}, \email{md.ashraful.alam1@g.bracu.ac.bd}, \email{iqbal.hossain@bracu.ac.bd}, \email{swakkhar.shatabda@bracu.ac.bd}}
\maketitle
\begin{abstract}
As network scale and uncertainty complicate automated penetration testing, reinforcement learning (RL) environments must support controlled evaluation under defensive and non-stationary conditions. We present DYNAS (Dynamic Network Attack Simulator), which combines parameterized large-scale scenario generation with stateful intrusion detection responses, explicit scan false-positive and false-negative noise, temporary service unavailability, and duration-based action timeouts. We further introduce Pen-DHRL, a recurrent manager--worker policy that selects persistent subgoals using IDS-conditioned global context. We compare Pen-DHRL with GNN, Invariant, and Attention baselines across four scenario variants and multiple training regimes. Under IDS-disabled training followed by IDS-enabled evaluation, Pen-DHRL obtains approximately 64\% and 24\% higher reward per step on Bridge and Dynamic, respectively, and approximately 26\% and 30\% more captured hosts on VarA and VarB than the next-best baseline. With all defensive dynamics active from the beginning of training, Pen-DHRL achieves positive mean episode returns in all four scenarios, whereas the flat baselines produce negative mean returns in all 12 architecture--scenario combinations; staged curriculum training produces positive mean returns across all 16 evaluated combinations.

\end{abstract}
\section{Introduction}
Automated security assessment provides repeatability and broad coverage, but automatically generated findings can require expert validation, whereas fully manual penetration testing is resource-intensive and difficult to scale \cite{rane2024comparative,nist2008testing}. This complementarity motivates learning-based methods for sequential attack planning, but their evaluation depends strongly on the assumptions represented by the training environment. Established RL environments such as NASim and NASimEmu provide efficient interaction, partially observable attack planning, and scenario variation, yet their published simulation protocols simplify stateful defensive reactions and several sources of observation and execution uncertainty \cite{schwartz2019nasim,janisch2023nasimemu}. DYNAS targets this evaluation gap through a controlled, defense-aware simulator for large parameterized network scenarios.\footnote{The DYNAS implementation is publicly available at \url{https://github.com/G5-Security/DYNAS}.}

NASimEmu was selected as the implementation base because its variable-size graph observations, parameterized scenario representation, and released RL baselines align directly with our questions concerning structural transfer and hierarchical control. Its common agent interface also permits policies to be evaluated across changing network instances while preserving a comparable action vocabulary. Together, these properties provide a technically suitable and reproducible foundation for DYNAS \cite{janisch2023nasimemu}.

\begin{table}
\caption{Abstraction-level and environmental-feature comparison of RL-based pentesting frameworks}\label{tab:framework_comparison}
\centering
\small
\setlength{\tabcolsep}{2pt}
\begin{tabular}{>{\raggedright\hyphenpenalty=10000\exhyphenpenalty=10000}p{2.9cm}ccccccc}
\hline
Feature & NASim & IAPTF & WCE & DynPen & NASimEmu & PenGym & DYNAS \\
\hline
Simulation-Based Training & \checkmark & \checkmark & Conceptual & \checkmark & \checkmark & \checkmark & \checkmark \\
Structural Scenario Variation & Limited & -- & Conceptual & \checkmark & Limited & Limited & \checkmark \\
Firewall-Aware Reachability & \checkmark & Limited & Conceptual & -- & \checkmark & \checkmark & \checkmark \\
IDS Representation/\allowbreak Response & -- & Limited & -- & -- & -- & -- & \checkmark \\
Curriculum Scheduling & -- & -- & -- & -- & -- & -- & \checkmark \\
Explicit Scan FP/FN Noise & -- & -- & -- & -- & -- & -- & \checkmark \\
Service-State Dynamics & -- & -- & -- & Limited & -- & -- & \checkmark \\
Duration-Based Action Timeouts & -- & -- & -- & -- & -- & -- & \checkmark \\
\hline
\end{tabular}
\end{table}
{\footnotesize Per NASim \cite{schwartz2019nasim}, IAPTF \cite{ghanem2023hierarchical}, WCE \cite{cody2022continuous,cody2023whole}, DynPen \cite{li2024dynpen}, NASimEmu \cite{janisch2023nasimemu}, and PenGym \cite{nguyen2024pengym}. $\checkmark$ denotes explicit support in the cited implementation; Limited denotes related support at a different abstraction or granularity; Conceptual denotes a proposed framework or NASim-based case study; and -- indicates that the feature is not documented in the cited protocol.}

NASimEmu already supports scenario randomization for evaluating transfer across network sizes and configurations. DYNAS extends this simulation protocol by jointly introducing stateful IDS responses, explicit scan errors, temporary service-state changes, duration-based action timeouts, and curriculum-controlled difficulty. The reviewed systems provide complementary subsets of these capabilities, but their published protocols do not document their joint use within one repeatable attack-focused simulator. Pen-DHRL is designed for this setting by combining recurrent global context, persistent high-level subgoals, and a goal-conditioned worker policy. Against this background, the primary contributions of this research are as follows:

\begin{itemize}\setlength{\itemsep}{1pt}\setlength{\parsep}{0pt}\setlength{\topsep}{3pt}
\item \textbf{Controlled Defense-Dynamics Layer:} A configurable simulation layer that jointly represents stateful IDS responses, explicit scan false-positive and false-negative noise, temporary service unavailability and recovery, and duration-based action timeouts under firewall-constrained reachability.
\item \textbf{Parameterized Scenario Generation:} A template-based generator that varies subnet sizes, host configurations, sensitive-host placement, and topology while preserving a stable action vocabulary.
\item \textbf{Hierarchical Policy:} A recurrent manager uses IDS context to select persistent subgoals that guide attack planning over long horizons.
\item \textbf{Regime-Specific Evaluation:} A comparison with flat RL baselines under defense-disabled training, immediate defended training, curriculum training, and transfer from IDS-disabled training to IDS-enabled evaluation.
\end{itemize}

\section{Motivation and Background}
\label{sec:motivation}

DYNAS and Pen-DHRL are shaped by three requirements.

R1: defensive reaction must be stateful (memory across steps, not a per-action penalty or scripted opponent) and configurable. DYNAS abstracts reaction into four mechanisms: stateful intrusion detection, triaged into isolation, patching, or heavier monitoring; scan false positives/negatives that corrupt observations; service churn, which can vanish a service on return; and action timeouts, where failure need not mean error. Sect.~\ref{sec:framework} details this layer.

R2: the agent must observe its own footprint.\label{subsec:threat} A penalty with no observable correlate makes stealth unlearnable, not merely difficult. DYNAS models an assumed-breach engagement and exposes each host's accumulated detection level, alerting threshold, and monitoring multiplier, so the agent can read its exact distance to a response rather than only a vague sense of noise; authorized engagements routinely grant this level of access, precisely so that an assessment can measure detection as well as compromise. The same reasoning gates host-internal actions, which require prior user-level access rather than remote capability. This motivates the IDS-conditioned context of Pen-DHRL (Sect.~\ref{subsec:pendhrl}).

R3: exposure to defence must be staged. With all four mechanisms active from episode one, a randomly initialized policy accumulates heavy penalty before reaching any sensitive host, favouring early termination over exploration---the regime in which every flat baseline we evaluate records negative reward per step, recovering only once training is staged (Table~\ref{tab:ids-curriculum-return}). This motivates the three-stage curriculum of Sect.~\ref{subsec:curriculum}, ordered easy to hard \cite{bengio2009curriculum}.

DYNAS addresses R1 and R3; Pen-DHRL is the policy designed for R2.

Preliminaries. Penetration testing is a partially observable Markov decision process: the observation and action set grow as subnets are discovered \cite{schwartz2019nasim}, motivating the graph-structured, permutation-equivariant policies used throughout \cite{janisch2023nasimemu}. Credit assigned over a single timescale dilutes across a long trajectory, since a sensitive host pays off only after discovery, exploitation, and privilege escalation all succeed; we address this with the feudal formulation \cite{dayan1992feudal,vezhnevets2017feudal}, where a manager emits a subgoal at a coarse timescale and a worker conditions on it. Subgoals may be semantic or latent; Pen-DHRL adopts the latter, so the decomposition is discovered from reward rather than imposed by hand.

\section{Related Work}
\label{sec:related}
Existing RL-based penetration-testing environments emphasize complementary research problems. NASim introduced an efficient simulator for learning attack paths but identified scalability and environment-fidelity as continuing challenges \cite{schwartz2019nasim}. IAPTF applied hierarchical decomposition to improve planning over large networks and represented security-aware clustering involving firewalls and IDSs \cite{ghanem2023hierarchical}. WCE proposed a broader continuous cyber-testing research framework and illustrated it through a NASim-based case study \cite{cody2022continuous,cody2023whole}. DynPen addressed changing host attributes and network connectivity \cite{li2024dynpen}, while NASimEmu emphasized scenario variation and transfer to unseen network configurations \cite{janisch2023nasimemu}. PenGym subsequently demonstrated the importance of testing learned policies under system-level action execution \cite{nguyen2024pengym}. Complementary efforts include automated red-team and post-exploitation methods \cite{filiol2021method,pham2024raiju}, deep reinforcement learning for IoT security \cite{samad8advancements}, and analyses of the shift from manual to automated testing \cite{kongara2023process,moreno2025analysis}; broader studies further motivate integrating learning-based and automated approaches \cite{alhamed2023systematic,ashawa2023penetration,badarneh2024power}.

A parallel line of work builds environments for autonomous cyber operations more broadly. CyberBattleSim models lateral movement, typically over a fixed, hand-specified topology, and includes a defender that detects and evicts the attacker \cite{microsoft2021cyberbattlesim}, acting as a scripted opponent rather than a parameterized detection layer. CybORG provides a gym for autonomous cyber agents \cite{standen2021cyborg} and underpins the CAGE challenges \cite{cage2022challenge2}, which are oriented toward training defensive agents. DYNAS is attack-focused, like CyberBattleSim, but exposes its defensive mechanisms as configurable, curriculum-schedulable parameters rather than a fixed opponent.

A further line of work evaluates policies on real infrastructure. Cyber ranges such as CyRIS \cite{pham2016cyris} and KYPO \cite{vykopal2017kypo} instantiate real hosts for security training. Such testbeds offer execution fidelity, while simulation offers the large, repeatable interaction budget and independent control over defensive mechanisms that a controlled evaluation framework requires. DYNAS is positioned deliberately on the simulation side.

Within learning-based attack planning, Hu et al.\ combine network-structure analysis with a Deep Q-Network to derive attack paths \cite{hu2020automated}, and Bu et al.\ apply a dual-stage DRL model to multi-subnet environments \cite{bu2025automated}. Closest to this work, Li et al.\ incorporate expert prior knowledge into a hierarchical DRL model for penetration testing \cite{li2023hierarchical}; the underlying manager--worker feudal formulation was established by Vezhnevets et al.\ \cite{vezhnevets2017feudal}. Pen-DHRL instead conditions its manager on an accumulated, per-host detection signal, so subgoal selection responds to the network's defensive state rather than to attacker-side progress alone.

These frameworks address distinct aspects of scalability, scenario variation, execution, and defensive uncertainty; Table~\ref{tab:framework_comparison} summarizes their complementary coverage. DYNAS integrates these capabilities within one controlled evaluation framework, while Pen-DHRL provides the corresponding hierarchical policy.

\section{Framework and Methodology}
\label{sec:framework}
\begin{figure}
\centering
\makebox[\textwidth][c]{\includegraphics[width=1.12\textwidth]{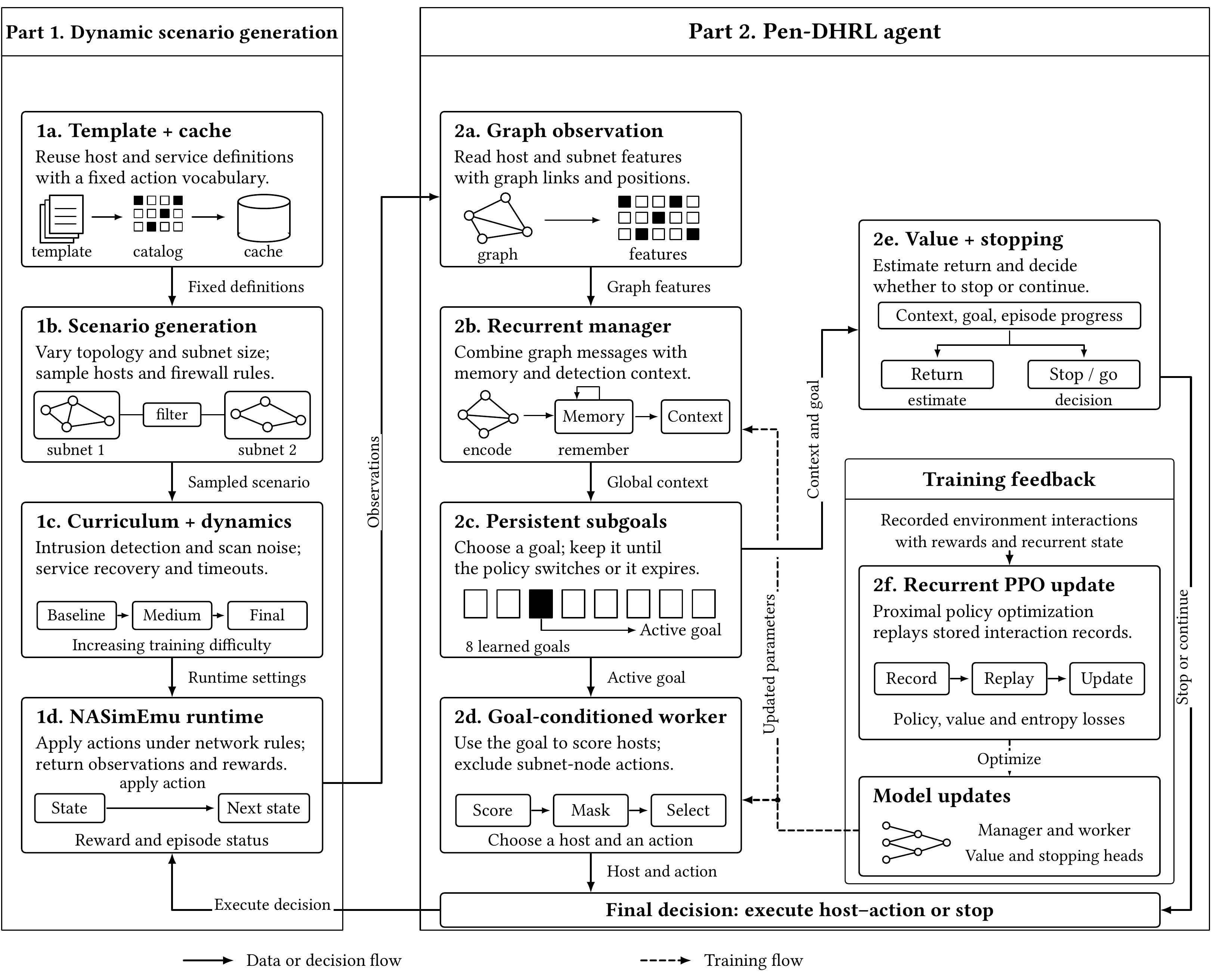}}
\Description{Two-part architecture: Part 1 generates dynamic scenarios from a cached template, scenario generation, curriculum injection, and the NASimEmu runtime. Part 2 contains the hierarchical agent pipeline, a value and stopping branch, and a separately bordered training-feedback section. Labeled solid arrows carry data and decisions; dashed arrows carry training flow. The final decision combines the worker's host-action choice with the stopping decision before returning to the environment. Unnecessary mathematical symbols are replaced with plain-language labels, and PPO is expanded as proximal policy optimization.}
\caption{Overview of Dynamic Scenario Generation and HRL Agent Development. Part~1 (left, detailed in Part 1 below) is the scenario pipeline, from the cached template through the NASimEmu runtime. Part~2 (right, detailed in Part 2 below) is the Pen-DHRL agent, from graph encoding through the value and termination heads. Solid arrows carry data and decisions during an episode; dashed arrows carry gradients during the recurrent PPO update.} \label{fig:overview-archi}
\end{figure}
DYNAS extends NASimEmu's simulation layer to generate larger parameterized scenarios and expose agents to controlled defensive and non-stationary effects.
Our contribution extends the \emph{simulation} side of NASimEmu. As shown in Fig.~\ref{fig:overview-archi}, the methodology has two connected parts.

\paragraph{Part 1: Dynamic Scenario Generation (1a--1d).}
\textbf{(1a)} A template YAML defines fixed primitives (OS, services, processes, exploits, and privilege-escalation rules) and is cached.
\textbf{(1b)} An auto-generator samples subnet sizes, topology style, host configurations, sensitive-host placement, and firewall constraints to create large dynamic scenarios.
\textbf{(1c)} Environment-dynamics parameters are configured through scenario or curriculum settings (when enabled), including IDS responses, scan noise, temporary service-state changes, and duration-based action timeouts.
\textbf{(1d)} The NASimEmu runtime executes state transitions, rewards, and constraints under these settings.

\paragraph{Part 2: Pen-DHRL Agent Pipeline (2a--2f).}
\textbf{(2a)} The agent receives graph observations with node/subnet features and positional encoding.
\textbf{(2b)} A recurrent manager GNN builds global context.
\textbf{(2c)} The manager selects persistent subgoals from a goal bank via a learned switch policy.
\textbf{(2d)} A goal-conditioned worker outputs node-action decisions.
\textbf{(2e)} Value and termination heads estimate return and stopping.
\textbf{(2f)} Parameters are optimized with recurrent PPO replay, and updates are fed back to both manager and worker.

We evaluate policy behavior across dynamically generated training and test instances and across the four defined scenario configurations using reward per step, episode return, episode length, and captured-host metrics. The experiments include defense-disabled, IDS-enabled curriculum, and defended-transfer conditions.

\subsection{Dynamic Scenario Generation}
The dynamic scenario generator of DYNAS extends the simulation layer of NASimEmu by combining a cached template layer with a parameterized auto-generator and a curriculum-controlled environment-dynamics stage. The template layer preserves a fixed action signature and observation dimensions across all generated instances, while the auto-generator resamples structural and parametric components per episode or per epoch.

\subsubsection{Extension to Dynamic Scenarios}
DYNAS generates scenarios from a YAML template that caches fixed primitives, including OS types, service and process probability maps, exploit definitions, and privilege escalation rules. This preserves an identical action signature across instances. The generator samples subnet counts and sizes from configurable ranges, selects a topology style (mesh, chain, or random), and instantiates each host by drawing its OS, services, and processes from the template probability maps. Sensitive host placement follows configurable host probabilities. Firewall rules are derived from the chosen connectivity style, and the curriculum or scenario settings configure IDS responses, scan noise, temporary service changes, and action timeouts at runtime.

\subsubsection{Domain Randomization}
To diversify training conditions, the pipeline independently varies exploit and privilege escalation success probabilities, action costs, and scan costs. Feature dropout additionally zeros observed service and process bits to simulate partial observability during reconnaissance. Template caching and action signature verification keep the exploit and privilege escalation space consistent across generated scenarios, preventing mismatches during batched rollouts. Curriculum injection activates environment parameters by stage.

\subsubsection{Curriculum Schedule}
\label{subsec:curriculum}
Training uses three curriculum stages defined over the training horizon. The baseline stage covers the first 30\% of training and disables IDS responses, scan noise, service churn, and action timeouts. The medium stage covers the next 30\% and jointly introduces low-intensity settings for all four mechanisms. The final stage covers the remaining 40\% and increases their configured difficulty. In the final stage, scan false-positive and false-negative rates range from 1.5\% to 7\% depending on scan type, the timeout probability is 1.5\%, the service-churn probability is 1\%, and IDS thresholds are sampled between 0.65 and 0.88. Evaluation uses the final stage, providing a consistent high-difficulty condition for policy comparison. The staged parameterization supports reproducible analysis of policy adaptation as defensive pressure and execution uncertainty increase.

\subsection{Pen-DHRL Model Design}
\label{subsec:pendhrl}
\begin{figure}[t]
\centering
\includegraphics[width=\textwidth]{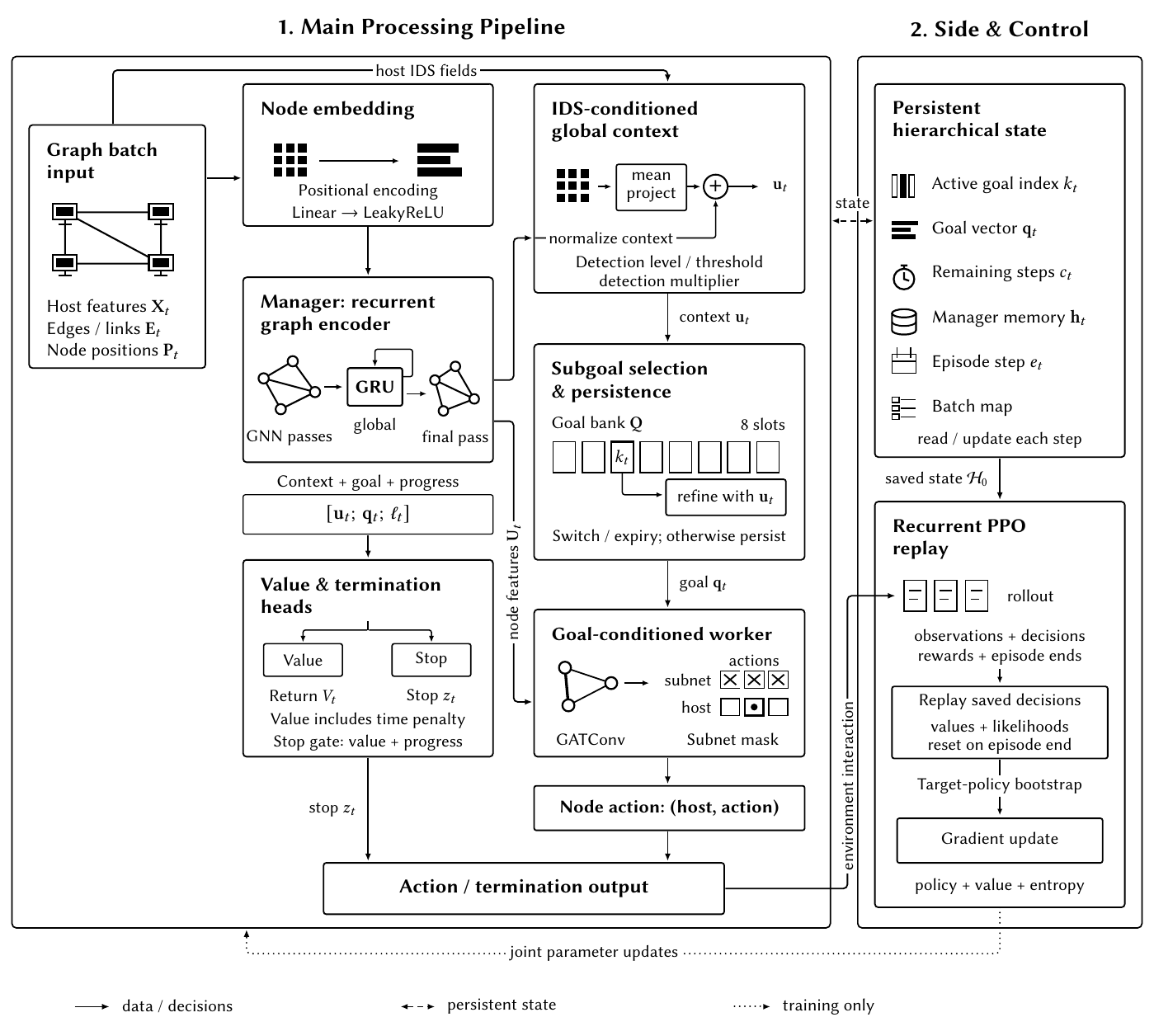}
\Description{Pen-DHRL architecture with a main processing pipeline from graph-batch input through node embedding, a recurrent manager graph encoder, IDS-conditioned global context, persistent subgoal selection, a goal-conditioned worker policy, and value and termination heads, plus side control blocks for persistent hierarchical state and recurrent PPO replay.}
\caption{Pen-DHRL Architecture Implemented in DYNAS. Left: the forward path for one decision. Node features and positional encodings are embedded and passed through the recurrent manager GNN to obtain per-node features and a graph-level global context; the graph-wise mean of the per-host IDS fields is projected and added to that context. The manager either keeps its active goal or draws a new slot from the goal bank, refines it with the global context, and passes the result to a GAT worker that emits a (host, action) decision. Value and termination heads read the concatenated global context, refined goal, and normalized episode progress. Right: the state carried across steps, and the recurrent PPO update, which snapshots that state, replays the recorded decisions with done-masked resets, and applies the clipped policy, value, and entropy losses.} \label{fig:pendhrl-arch}
\end{figure}
As shown in Fig.~\ref{fig:pendhrl-arch}, Pen-DHRL is implemented as a simulation-first pipeline in DYNAS, combining controlled dynamic scenarios with hierarchical graph control. Relative to the original NASimEmu, the DYNAS simulation layer configures IDS responses, scan noise, temporary service-state changes, and duration-based action timeouts through scenario or curriculum settings, while the agent layer performs recurrent manager-worker decision making over graph observations.

\subsubsection{Main Processing Pipeline}
Given a graph-observation batch, the recurrent manager produces node features and global context, to which it adds the graph-wise mean of the host IDS fields observable to the agent: the accumulated detection level, alerting threshold, and monitoring multiplier (Sect.~\ref{subsec:threat}). The baseline represents its subgoals as \(K=8\) learned latent slots in a goal bank \(\mathbf{Q}\in\mathbb{R}^{8\times128}\). A learned switch and a counter initialized to \(H=4\) govern slot refresh, after which the selected goal is refined by the current context and supplied to the worker.

\subsubsection{Side and Control Mechanisms}
Pen-DHRL persists the latent-slot index and vector, counter, manager hidden state, batch map, and episode counters. Reset, masked-reset, and clone operations support recurrent training. The recurrent PPO update replays the recorded worker action, stopping decision, and active goal index from the initial state with done masks, forms bootstrapped returns, and jointly updates both controllers.

\subsection{Training and Evaluation Protocol}
\label{sec:training-evaluation-protocol}
For the defended comparisons, all policies were optimized using a common vectorized PPO training loop. Each optimization iteration collected an eight-step rollout from 128 parallel environments, yielding 1{,}024 transitions, followed by three clipped PPO passes. The full training schedule comprised 200 reporting intervals of 100 PPO iterations, corresponding to 20{,}000 optimization iterations. PPO used a discount factor of \(0.99\), a clipping parameter of \(0.2\), and gradient-norm clipping at \(3.0\). Parameters were optimized using AdamW with an initial learning rate of \(7\times10^{-4}\), weight decay of \(10^{-4}\), and an initial entropy coefficient of \(0.02\). Bootstrap targets were computed using a target network updated by soft parameter averaging with coefficient \(\rho=0.1\). Pen-DHRL additionally preserved its recurrent manager state during rollout and replay, as described in the preceding model-design subsection.

At each episode reset, a scenario was selected uniformly from the configured training pool and regenerated by resampling subnet sizes, sensitive-host placement, and host OS, service, and process configurations while preserving a common action vocabulary. Curriculum difficulty was updated once per reporting interval. Evaluation was performed without parameter updates using 64 fresh parallel environments, with recurrent state reset at episode boundaries. Curriculum-enabled evaluation environments always used the final configured defense-dynamics stage. Each evaluation pass used a target of 100 completed episodes and a 400-step interaction limit. Episodes terminated when the policy selected its terminal action or reached this limit, and action generation followed each architecture's implemented inference rule. Episode-level metrics were normalized by the number of completed episodes, while reward per step was normalized by the number of executed interactions.

\section{Experimental Evaluation}
\subsection{Evaluation Metrics and Interaction Scale}
\subsubsection{Evaluation Metrics}
Rewards follow the implemented scaled transition rule $r=(v-c)/10$, using scenario-defined host values and action costs. We evaluate each policy using four complementary metrics: reward per step measures mean utility per interaction, mean episode return measures cumulative utility, episode length reflects interaction efficiency, and the mean number of distinct sensitive hosts captured measures attack coverage separately from return. All metrics follow the protocol in Sect.~\ref{sec:training-evaluation-protocol}.

Here $v$ is non-zero only when an action first raises access to root on a sensitive host or when an intrusion-detection response fires, and $c$ is the scenario-defined cost of the executed action. The divisor of ten is a fixed global scaling, applied to all reported returns and penalties.

\subsubsection{Interaction-Budget Reference}
To characterize the interaction scale of each scenario, consider a complete-network sweep requiring one scan per subnet, two reconnaissance actions per host, one process scan per sensitive host, and expected exploit and privilege-escalation retries. Assuming success probabilities of approximately $0.8$, the resulting heuristic action budget is
\begin{equation}
    A_{\text{heuristic}} = S + 2H + 3.5N.
    \label{eq:action-budget}
\end{equation}
Here, $S$, $H$, and $N$ denote the numbers of subnets, hosts, and sensitive hosts, respectively. The heuristic combines subnet discovery, host reconnaissance, and expected compromise attempts into a common interaction-scale reference across scenarios.

Table~\ref{tab:train-test} lists the scenario dimensions and heuristic action budget alongside the observed capture and episode-return values in the defense-disabled setting.

\begin{table}
\caption{Scenario dimensions, heuristic interaction budget, and observed Pen-DHRL outcomes in the defense-disabled setting}\label{tab:train-test}
\centering
\small
\begin{tabular}{lcccc}
\hline
Scenario & $(N,H,S)$ & $A_{\text{heuristic}}$ & Captured hosts & Episode return \\
\hline
Corp-Dyn & $(39,83,12)$ & $314.5$ & $36.13$ & $49.41$ \\
VarA     & $(41,86,12)$ & $327.5$ & $36.68$ & $48.43$ \\
\hline
\end{tabular}
\end{table}

\subsection{Scenario Characterization}
\subsubsection{Reference and Dynamic Scenarios}
Tables~\ref{tab:scenario-services} and~\ref{tab:scenario-structure} contrast original NASimEmu configurations with DYNAS dynamic configurations. The original configurations are \texttt{corp} and \texttt{uni}; the dynamic configurations are \texttt{corp\_dynamic} and \texttt{corp\_dynamic\_varA}. The reference configurations establish the original scenario scale, while the dynamic configurations introduce larger network instances and broader host, service, process, and exploit catalogs.
\begin{itemize}
\item \textbf{corp.v2:} Enterprise baseline with 8 subnets, 21 hosts, and 6 services.
\item \textbf{uni.v2:} Academic-like topology with 9 subnets, 27 hosts, and 6 services.
\item \textbf{corp\_dynamic.v2:} Expanded network (12 subnets, 83 hosts) with 32 services, 13 processes, and 8 exploits across Linux/Windows stacks.
\item \textbf{corp\_dynamic\_varA:} Variant of \texttt{corp\_dynamic} with lower exploit probabilities and higher costs, producing stronger stochastic behavior.
\end{itemize}

\paragraph{Service, Process, and Structural Comparison.}
Relative to the reference configurations, the dynamic scenarios contain more hosts, services, processes, exploits, and sensitive hosts (Tables~\ref{tab:scenario-services} and~\ref{tab:scenario-structure}), increasing both graph size and the range of attack-relevant configurations encountered during policy learning.

\begin{table}
\caption{Service and process distribution across network scenarios}\label{tab:scenario-services}
\centering
\small
\setlength{\tabcolsep}{2pt}
\begin{tabular}{lccccccccc}
\hline
Scenario & LinSvc & WinSvc & AnySvc & SensSvc & LinProc & WinProc & UserExp & RootExp & PrivEsc \\
\hline
Corp     & 3  & 2 & 1 & 1 & 0 & 0 & 4 & 1 & 1 \\
Uni      & 3  & 2 & 1 & 1 & 0 & 0 & 4 & 1 & 1 \\
Corp-Dyn & 23 & 8 & 1 & 3 & 6 & 7 & 7 & 1 & 6 \\
VarA     & 23 & 8 & 1 & 3 & 6 & 7 & 8 & 0 & 6 \\
\hline
\end{tabular}
\end{table}

\begin{table}
\caption{Structural and security characteristics of network scenarios}\label{tab:scenario-structure}
\centering
\small
\begin{tabular}{lccccc}
\hline
Scenario & Subnets & Hosts & Exploits & SensHosts & Density \\
\hline
Corp     & 8  & 21 & 5 & 7  & 0.25 \\
Uni      & 9  & 27 & 5 & 3  & 0.27 \\
Corp-Dyn & 12 & 83 & 8 & 39 & 0.19 \\
VarA     & 12 & 86 & 8 & 41 & 0.19 \\
\hline
\end{tabular}
\end{table}

\subsection{RL Architecture Comparison}
\subsubsection{Dynamic Baseline}
The original NASimEmu baselines include GNN, Attention, and Invariant agents, which provide useful comparisons for flat policy learning across training and test scenarios. With defenses disabled, Table~\ref{tab:final} shows that Pen-DHRL obtains the highest reward per step and mean episode return. GNN achieves slightly greater host capture, while Invariant produces shorter episodes. The results therefore indicate improved return under hierarchical control while retaining a measurable efficiency and coverage trade-off.

\begin{table}
\caption{Baseline performance on dynamically generated scenarios without IDS or curriculum}\label{tab:final}
\centering
\small
\begin{tabular}{llrr}
\hline
Metric & Architecture & Training & Testing \\
\hline
Reward/step & GNN       & 0.27 & 0.27 \\
            & Invariant & 0.34 & 0.33 \\
            & Attention & 0.26 & 0.26 \\
            & Pen-DHRL  & 0.49 & 0.48 \\
\hline
Episode return & GNN       & 29.90 & 30.02 \\
               & Invariant & 27.21 & 27.06 \\
               & Attention & 28.21 & 28.09 \\
               & Pen-DHRL  & 49.41 & 48.43 \\
\hline
Episode length & GNN       & 111.50 & 112.63 \\
               & Invariant &  81.10 &  80.96 \\
               & Attention & 109.90 & 109.42 \\
               & Pen-DHRL  & 103.40 & 106.40 \\
\hline
Captured hosts & GNN       & 36.91 & 37.10 \\
               & Invariant & 32.23 & 32.13 \\
               & Attention & 35.63 & 35.46 \\
               & Pen-DHRL  & 36.13 & 36.68 \\
\hline
\end{tabular}
\end{table}

\subsubsection{Defended Evaluation Scenarios}
The defended evaluation uses four complementary configurations with a shared action vocabulary. Dynamic provides the reference large-network setting; Bridge interpolates host, service, process, exploit, and action-cost parameters; VarA combines lower exploit probabilities, higher costs, and a redistributed sensitive-host profile; and VarB applies both a topology change and a further parameter shift. Together, these configurations form a structured progression from interpolation to joint structural--parametric transfer while retaining comparable action semantics.

\subsubsection{IDS-Enabled Training: Curriculum vs.\ No-Curriculum}
Tables~\ref{tab:ids-curriculum-return} and~\ref{tab:ids-curriculum-coverage} compare all four architectures under IDS-enabled evaluation on Dynamic, Bridge, VarA, and VarB, reporting each metric under \emph{no-curriculum} and \emph{curriculum} training together with the change $\Delta$ (curriculum minus no-curriculum). In training without a curriculum, all four mechanisms are active from the outset. In that regime, the flat baselines produce negative reward per step, capture few hosts, and frequently reach the step limit; Pen-DHRL remains positive. The curriculum comprises three stages: the baseline stage disables these mechanisms, the medium stage jointly introduces them at low intensity, and the final stage increases their configured difficulty. Across this comparison, each flat baseline moves to positive reward per step and gains approximately 9--28 captured hosts, while Pen-DHRL improves more modestly because its no-curriculum performance is already positive. Table~\ref{tab:ids-off-training} evaluates a distinct transfer regime in which IDS is disabled during training and enabled during evaluation. In this regime, Pen-DHRL obtains the highest reward per step on Bridge and Dynamic and the highest capture on VarA and VarB; Attention produces the shortest episodes on VarA and VarB but captures fewer than one host on average.

\begin{table}[htbp]
\caption{IDS-enabled evaluation of reward per step and episode return: no-curriculum vs.\ curriculum. Episode returns are rounded to integers; each $\Delta$ uses unrounded values}\label{tab:ids-curriculum-return}
\centering
\small
\setlength{\tabcolsep}{3.5pt}
\begin{tabular}{llrrrrrr}
\hline
& & \multicolumn{3}{c}{Reward/step} & \multicolumn{3}{c}{Episode return} \\
\cline{3-5}\cline{6-8}
Scenario & Model & No & Curr & $\Delta$ & No & Curr & $\Delta$ \\
\hline
Bridge
    & Pen-DHRL  & $\phantom{-}0.40$ & $1.51$ & $+1.11$ & $\phantom{-}161$ & $213$ & $+53$  \\
    & GNN       & $-0.08$           & $1.37$ & $+1.45$ & $-31$            & $189$ & $+220$ \\
    & Invariant & $-1.30$           & $1.22$ & $+2.52$ & $-511$           & $206$ & $+717$ \\
    & Attention & $-1.23$           & $1.24$ & $+2.47$ & $-490$           & $205$ & $+695$ \\
\hline
Dynamic
    & Pen-DHRL  & $\phantom{-}0.54$ & $2.25$ & $+1.71$ & $\phantom{-}217$ & $249$ & $+32$  \\
    & GNN       & $-0.08$           & $2.17$ & $+2.25$ & $-30$            & $254$ & $+284$ \\
    & Invariant & $-1.41$           & $1.89$ & $+3.30$ & $-560$           & $263$ & $+822$ \\
    & Attention & $-1.31$           & $2.01$ & $+3.32$ & $-523$           & $253$ & $+776$ \\
\hline
VarA
    & Pen-DHRL  & $\phantom{-}0.17$ & $0.61$ & $+0.44$ & $\phantom{-}68$  & $134$ & $+66$  \\
    & GNN       & $-0.10$           & $0.64$ & $+0.74$ & $-41$            & $139$ & $+180$ \\
    & Invariant & $-1.29$           & $0.63$ & $+1.92$ & $-514$           & $140$ & $+654$ \\
    & Attention & $-1.20$           & $0.67$ & $+1.87$ & $-479$           & $134$ & $+613$ \\
\hline
VarB
    & Pen-DHRL  & $\phantom{-}0.08$ & $0.38$ & $+0.30$ & $\phantom{-}31$  & $83$  & $+52$  \\
    & GNN       & $-0.06$           & $0.40$ & $+0.46$ & $-28$            & $88$  & $+115$ \\
    & Invariant & $-1.18$           & $0.40$ & $+1.58$ & $-458$           & $98$  & $+556$ \\
    & Attention & $-0.56$           & $0.19$ & $+0.75$ & $-224$           & $48$  & $+271$ \\
\hline
\end{tabular}
\end{table}

\begin{table}[htbp]
\caption{IDS-enabled evaluation of episode length and captured hosts: no-curriculum vs.\ curriculum. Episode lengths are rounded to integers; each $\Delta$ uses unrounded values}\label{tab:ids-curriculum-coverage}
\centering
\small
\setlength{\tabcolsep}{3.5pt}
\begin{tabular}{llrrrrrr}
\hline
& & \multicolumn{3}{c}{Episode length} & \multicolumn{3}{c}{Captured hosts} \\
\cline{3-5}\cline{6-8}
Scenario & Model & No & Curr & $\Delta$ & No & Curr & $\Delta$ \\
\hline
Bridge
    & Pen-DHRL  & 400 & 141 & $-259$ & 24.95 & 25.72 & $+0.77$  \\
    & GNN       & 400 & 138 & $-262$ &  0.99 & 23.68 & $+22.69$ \\
    & Invariant & 392 & 170 & $-222$ &  2.24 & 25.56 & $+23.32$ \\
    & Attention & 400 & 165 & $-235$ &  1.07 & 23.72 & $+22.65$ \\
\hline
Dynamic
    & Pen-DHRL  & 400 & 113 & $-287$ & 29.59 & 30.37 & $+0.78$  \\
    & GNN       & 400 & 117 & $-283$ &  1.00 & 28.49 & $+27.49$ \\
    & Invariant & 397 & 139 & $-258$ &  2.15 & 29.80 & $+27.65$ \\
    & Attention & 400 & 126 & $-274$ &  1.06 & 27.44 & $+26.38$ \\
\hline
VarA
    & Pen-DHRL  & 400 & 238 & $-162$ & 13.73 & 19.62 & $+5.89$  \\
    & GNN       & 400 & 218 & $-182$ &  0.00 & 18.46 & $+18.46$ \\
    & Invariant & 397 & 223 & $-175$ &  0.00 & 18.13 & $+18.13$ \\
    & Attention & 400 & 200 & $-200$ &  0.00 & 17.29 & $+17.29$ \\
\hline
VarB
    & Pen-DHRL  & 400 & 249 & $-151$ &  9.39 & 14.93 & $+5.54$  \\
    & GNN       & 400 & 220 & $-180$ &  0.00 & 13.83 & $+13.83$ \\
    & Invariant & 387 & 243 & $-144$ &  0.00 & 14.52 & $+14.52$ \\
    & Attention & 400 & 255 & $-145$ &  0.00 &  9.48 & $+9.48$  \\
\hline
\end{tabular}
\end{table}

\begin{table}[htbp]
\caption{Per-scenario model performance with IDS disabled during training but enabled during evaluation}\label{tab:ids-off-training}
\centering
\small
\begin{tabular}{llrrrr}
\hline
Scenario & Architecture & Reward/step & Episode return & Ep Len & Capture \\
\hline
Bridge
    & Pen-DHRL  & $\phantom{-}1.15$ & $\phantom{-}161.96$ & 141.20 & 23.36 \\
    & GNN       & $\phantom{-}0.18$ & $\phantom{-}66.20$  & 368.86 & 26.17 \\
    & Invariant & $\phantom{-}0.01$ & $\phantom{-}5.20$   & 367.77 & 25.69 \\
    & Attention & $\phantom{-}0.70$ & $\phantom{-}120.04$ & 170.39 & 17.58 \\
\hline
Dynamic
    & Pen-DHRL  & $\phantom{-}1.49$ & $\phantom{-}200.17$ & 134.56 & 27.80 \\
    & GNN       & $\phantom{-}0.26$ & $\phantom{-}91.32$  & 356.31 & 30.59 \\
    & Invariant & $\phantom{-}0.19$ & $\phantom{-}66.58$  & 348.37 & 30.88 \\
    & Attention & $\phantom{-}1.20$ & $\phantom{-}204.54$ & 170.28 & 26.87 \\
\hline
VarA
    & Pen-DHRL  & $\phantom{-}0.14$ & $\phantom{-}46.42$ & 323.19 & 16.53 \\
    & GNN       & $\phantom{-}0.38$ & $\phantom{-}62.45$ & 162.44 & 10.53 \\
    & Invariant & $\phantom{-}0.14$ & $\phantom{-}46.08$ & 327.06 & 13.17 \\
    & Attention & $-0.11$           & $-8.50$            &  74.20 &  0.34 \\
\hline
VarB
    & Pen-DHRL  & $\phantom{-}0.11$ & $\phantom{-}28.66$ & 266.58 & 11.42 \\
    & GNN       & $\phantom{-}0.22$ & $\phantom{-}36.52$ & 169.74 &  8.26 \\
    & Invariant & $\phantom{-}0.08$ & $\phantom{-}19.89$ & 264.46 &  8.79 \\
    & Attention & $-0.13$           & $-5.24$            &  39.64 &  0.10 \\
\hline
\end{tabular}
\end{table}

The $\Delta$ columns report the curriculum-associated change for each metric. Every flat baseline's reward per step changes from negative to positive under the curriculum; for example, Invariant on Dynamic increases by $+3.30$, from $-1.41$ to $1.89$, whereas Pen-DHRL, already positive, records smaller gains. Episode-return changes reach $+822$, and episode length decreases by approximately 140--290 steps for the flat baselines. Captured-host counts increase by approximately 9--28 for the flat baselines, while the corresponding Pen-DHRL changes remain below six.

\section{Discussion}
The results show that the training regime materially changes model ranking. When all defensive dynamics are active from the beginning of training, Pen-DHRL remains positive across all four scenarios, while the flat baselines obtain negative returns and very low capture. Curriculum scheduling improves every architecture and makes the flat baselines competitive, indicating that progressive exposure is particularly important for non-hierarchical policies.

With curriculum training, Pen-DHRL obtains the highest reward per step on Bridge and Dynamic by approximately $10\%$ and $4\%$, respectively. On VarA and VarB, Pen-DHRL retains the highest capture, while the flat models obtain slightly higher reward per step. Under IDS-disabled training followed by IDS-enabled evaluation, Pen-DHRL's advantage is larger: it provides approximately $64\%$ and $24\%$ higher reward per step on Bridge and Dynamic and approximately $26\%$ and $30\%$ greater capture on VarA and VarB. Together, these results identify defended transfer as the regime in which Pen-DHRL provides its strongest reward--coverage profile.

The evaluation assesses Pen-DHRL as an integrated hierarchical architecture in which recurrent global context, IDS conditioning, and persistent subgoals jointly coordinate long-horizon decisions. The resulting policy behaviour is consistent with the stronger reward--coverage profile observed under defended transfer.

\section{Conclusion and Future Work}
This work presents DYNAS, a Dynamic Network Attack Simulator that combines parameterized large-network generation with stateful IDS responses, explicit scan noise, temporary service-state changes, duration-based action timeouts, and curriculum-controlled difficulty. Within this environment, Pen-DHRL integrates recurrent global context, persistent high-level subgoals, and an IDS-conditioned worker policy. The evaluation shows that Pen-DHRL provides its clearest advantage when transferring from defense-disabled training to defended evaluation, while curriculum scheduling substantially improves all evaluated architectures and narrows the performance gap. These findings support hierarchical control as a strong approach for reward--coverage balance under defended and non-stationary conditions. Future work will extend DYNAS across additional topology families and integrate active defender policies and host-level firewall rules.

\bibliographystyle{splncs04}
\bibliography{refs}
\end{document}